# An Optical Reflector System for the CANGAROO-II Telescope

Akiko Kawachi for the CANGAROO Collaboration[1]

*Institute for Cosmic Ray Research, University of Tokyo Tanashi, Tokyo 188-8502, Japan*[2]

**Abstract.** We have developed light and durable mirrors made of CFRP (Carbon Fiber Reinforced Plastics) laminates for the reflector of the new CANGAROO-II 7 m telescope. The reflector has a parabolic shape (F/1.1) with a 30 $m^2$ effective area which consists of 60 small spherical mirrors. The attitude of each mirror can be remotely adjusted by stepping motors. After the first adjustment work, the reflector offers a point image of about $0^\circ.14$ (FWHM) on the optic axis. The telescope has been in operation since May 1999 with an energy threshold of ∼300 GeV.

## INTRODUCTION

The CANGAROO collaboration has started operation of a new imaging Cherenkov telescope of 7 m reflector. With an effective light collecting area of 30 $m^2$, it enables us to observe very high-energy gamma-rays of the southern sky with a 300 GeV energy threshold [1,2]. The telescope is as well a prototype of the project of an array of four 10 m telescopes, CANGAROO-III [3].

## MECHANICAL DESIGN

The reflector is an F/1.1 paraboloid with a diameter of 7 m. In order to use timing information to reject night sky background, we chose a parabolic shape which provides isochronous collection of photons. Sixty spherical mirrors, each of which has an 80 cm diameter and a curvature radius between 16–17 m, were arranged according to their curvature radii from the inner to the outer sections of the reflector, with the shorter focal length mirrors innermost. In the prime focal plane, there is a multi-pixel camera with $0.^\circ 12$ pitch covering about $3^\circ$ of FOV.

The design of the supporting frame of the reflector was based on a commercially available communications antenna, and the frame is mounted by 9 honeycomb panels. Several mirrors (6–9) were installed onto a honeycomb panel and the alignment

1) see [3] for a complete name list of the collaboration.
2) E-mail: kawachi@icrr.u-tokyo.ac.jp

of the mirrors in each panel was roughly adjusted ($\lesssim 0.^\circ 3$) with a laser beam before the shipping to Australia. With these adjustments, we checked our remote adjustment system as well as it saved on-site labor. The total weight of the dish including the mirrors and adjustment system is very light; only 4.3 ton (6.3 ton when the camera stays are included), so that we could greatly reduce cost and assembling labor. The structure was designed to be operated at the average velocity 30 km/hr, operational up to 100 km/hr of the wind load. By tracking various stars at 12–85 degrees in the elevation angle and at all the azimuthal angles, gravitational deformations of the structure were measured to cause less than $1'$ of deviation in the pointing accuracy at the focal plane

The present support frame allows us to extend the reflector up to 10 m diameter with additional 54 mirrors, and the extension is to be completed by early 2000.

## SMALL SPHERICAL MIRRORS

The small spherical mirror is of 80 cm in diameter, of about 2 cm thick, and weighs only 5.5 kg. Sheets of CFRP and adhesives were laid on a metal mold, sandwiching a core of low density, high shear strength foam to avoid twisting deformations, and a polymer sheet coated with laminated aluminum was applied on the top of the layers as a reflecting material. The laminates were vacuum bagged and cured to 120°C in an autoclave pressure vessel.

For protection against dust, rain, and sunshine, the mirror surface was coated with fluoride. The reflectivity is over 80 % at 340 nm–800 nm, falls off rather slowly down to 40 % at 250 nm. We found dusts on the surface deteriorate the reflectivity to about 75 % after several months, however, it was confirmed with a sample in a year-time-scale that the reflectivity repeatedly recovers easily by water washing. The mirror surface is free from dewing until the relative humidity exceeds 83 %.

The curvature radii of the mirrors show a flat distribution between 15.9–17.1 m, with an average of 16.45 m. The mirrors were arranged on the support according to their radii to make a smooth $f$=8 m paraboloid, and the individual facets were adjusted toward the focal point by the method described later. The image size of each mirror was measured with a light source 5.8 km away. A typical size is $0.^\circ 08$ (FWHM), and 50 % of the photons concentrates within $\sim 0.^\circ 1\ \phi$.

## REMOTE ADJUSTMENT SYSTEM

Two watertight stepping motors are installed at the back surface of each mirror, and the attitude of a mirror can be remotely adjusted in two perpendicular directions. The minimum step size corresponds to about $1\times10^{-4}$ degree at the focal plane, and adjustable up to $\pm$3-degree. The accuracy of $1\times10^{-3}$ degree is retained when motors are switched off. All mirrors are adjusted one by one using two motor drivers with relay switches controlled by a computer.

Our adjustment work on-site used a distant light source at night. All small mirrors were covered but one with plastic lids, and its image on a screen at the focal plane was monitored by a CCD camera installed at the center of the reflector. The attitudes of the mirrors were adjusted by moving stepping motors using feedback information from CCD images so that the image center should lay at the focal point.

As a result of the first adjustment work, the deviation of the small mirror axis orientations is $\sim 0^\circ.03$ on average, a larger value than expected. The deviation was mainly caused by temporal fluctuations of the CCD camera geometry over different nights we applied the adjustment. Removing this effect, it is estimated that the orientations can be adjusted within an error of $0^\circ.01$.

## PERFORMANCES OF THE REFLECTOR

The optical property of the reflector in total was measured using images of several stars tracked by the telescope. Images on the focal plane screen were taken by a CCD camera at the reflector center.

### A On the Optical Axis

In Figure 1, an image of Sirius on the optic axis is shown in units of CCD pixels. A pixel corresponds to $6.7\times10^{-3}$ degree. One pixel of the camera ($0^\circ.12$ square) is overdrawn for scaling. An image size of 0°.14±0°.01 (FWHM) is deduced, and 30±4 % of the photons concentrates in a single camera pixel. The characteristic difference in size between the images of gamma-rays and protons is on the scale of 0.°1–0.°2. Our optical quality meet the requirement to start with, though there is room for improvement. A preliminary analysis of the CANGAROO-II data shows thin muon rings whose average width is ~0.°11 [4]. Contribution of mirror aberrations to broadening the images of rings can be estimated as comparable to that of multiple scatterings in the atmosphere [5].

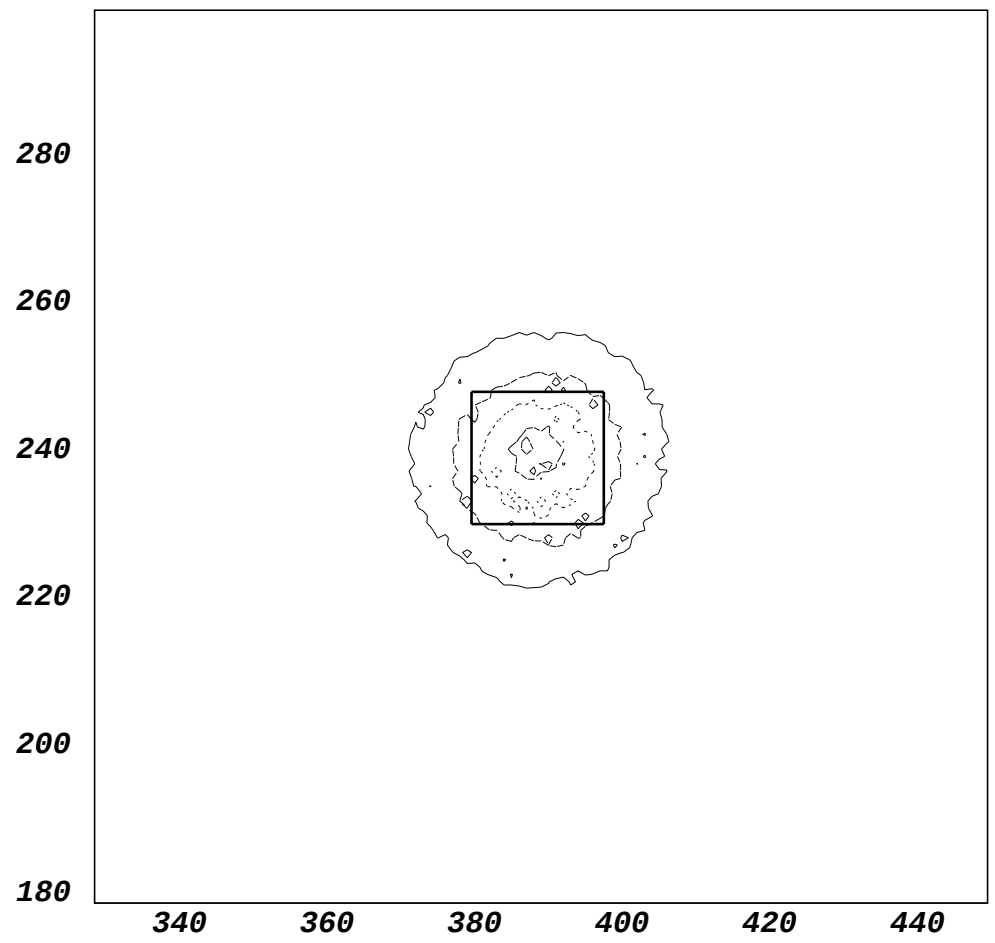

**FIGURE 1.** A CCD image of Sirius on the optic axis. The axes are in unit of CCD pixels, corresponding to a $6.7\times10^{-3}$ degree pitch. $z$ axis is in arbitrary unit. A square overdrawn is a scale of a camera pixel ($0^\circ.12$ square).

## B Off the Optical Axis

Parabolic mirrors have greater off-axis aberrations. The aberration is rather serious for a Cherenkov imaging telescope since a relatively wide field of view (~3°) is needed for image analyses of atmospheric showers. We compared off-axis images of Sirius displacing the pointing coordinates both in right ascension and in declination. As a result of symmetric configuration and alignments, displacement to all the directions caused deformations consistent with each other (Figure 2 *top*)). Figure 2 *bottom*) shows radial point spread functions of the Sirius pointed on and off in right ascension angles by $0.^\circ5$, $1.^\circ0$ and $1.^\circ3$, respectively, We can estimate concentration decrease by about 18 % at the edge of the FOV.

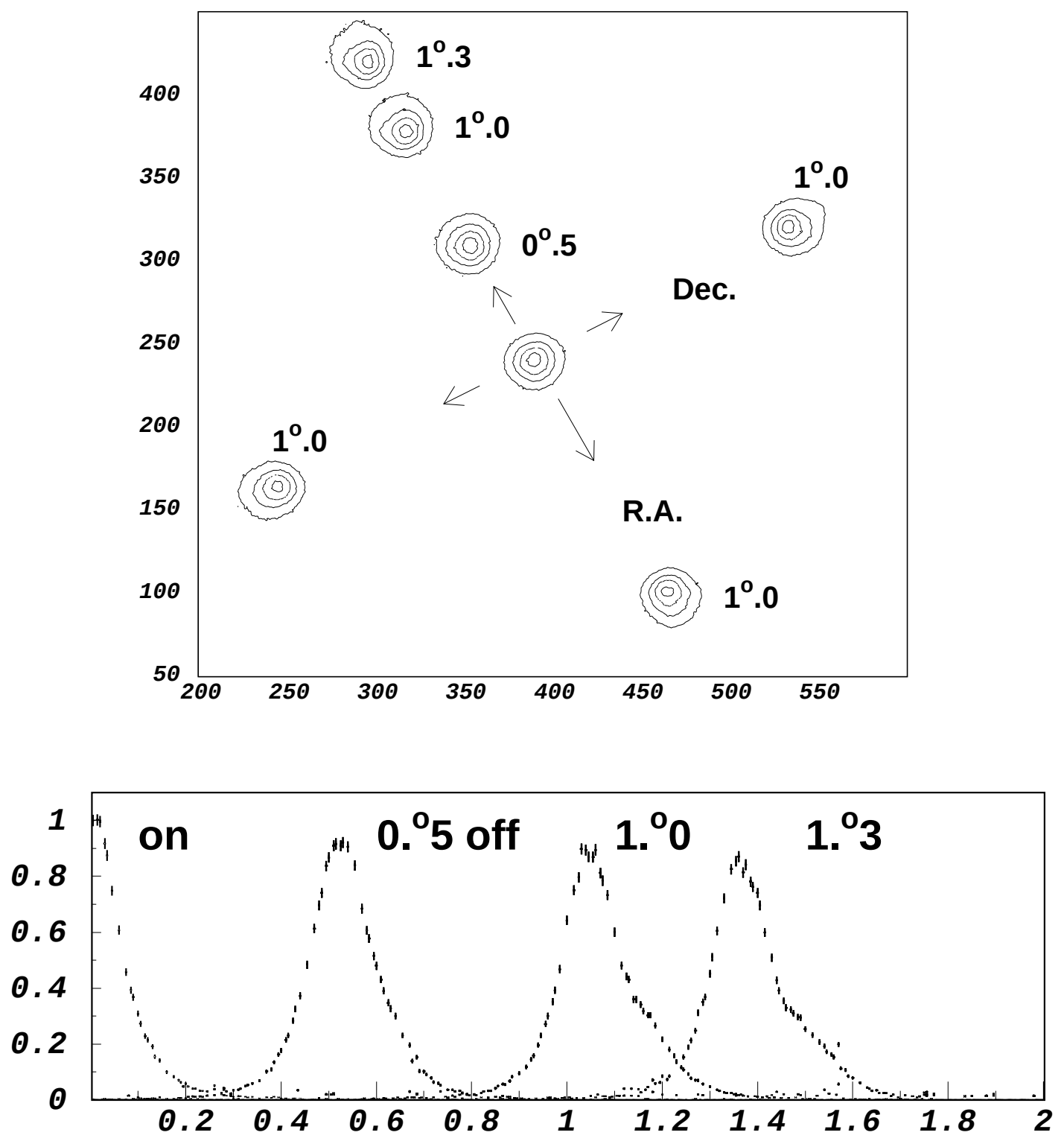

**FIGURE 2.** *top*) A synthesized figure of the CCD images obtained by displacing the pointing coordinates of a star. *bottom*) Radial point spread functions at the different pointing coordinates; ($\alpha$, $\delta$), ($\alpha$−0.°5, $\delta$) ($\alpha$−1.°0, $\delta$) and ($\alpha$−1.°3, $\delta$), respectively. Vertical scales are normalized by the peak height of the on-axis psf.

## C Gravitational Deformation

Since the adjustment work of the small mirrors was performed pointing horizontally, possible deflections of the alignment system at other attitudes had to be calibrated. The effect of gravitational deformations on the reflector was measured by comparing the images of stars taken at various elevation angles of the telescope. For elevation angles between 15–70 degrees, the images show no dependence on the elevations either in shape or in size. Thus the deformations are confirmed to be negligible within an error of $0.^\circ01$.

## SUMMARY

The new CANGAROO-II 7 m telescope has been completed and operations has begun. The reflector, F/1.1 paraboloid, has a point spread function of $0.^\circ14$ (FWHM) over 3 degrees of FOV, with $\sim$20 % loss of light concentration by aberration at the FOV edge.

## ACKNOWLEDGMENTS

The small mirrors of CFRP laminates have been developed in collaboration with Mitsubishi Electric Corporation, Communication Systems Center. AK was supported for this work by a JSPS postdoctoral fellowship.